# Reconfigurable microresonators induced in side-coupled optical fibers

**V. Vassiliev and M. Sumetsky***

*Aston Institute of Photonic Technologies, Aston University, Birmingham B4 7ET, UK*
*\*Email: m.sumetsky@aston.ac.uk*

**We experimentally demonstrate that side-coupling of coplanar bent optical fibers can induce a high Q-factor whispering gallery mode (WGM) optical microresonator. To explain the effect, we consider WGMs with wavelengths close to the cutoff wavelengths (CWs) of these fibers which slowly propagate along the fiber axes. In the vicinity of the touching region, WGMs of adjacent fibers are coupled to each other, and CWs experience sub-nanoscale axial variation proportional to the coupling strength. We show that in certain cases the CW variation leads to full localization of the WGMs and the creation of an optical microresonator. By varying the characteristic curvature fiber radius from the centimeter order to millimeter order, we demonstrate fully mechanically reconfigurable high Q-factor optical microresonators with dimensions varying from the millimeter order to 100-micron order and free spectral range varying from a picometer to hundreds of picometers. The new microresonators may find applications in cavity QED, microresonator optomechanics, frequency comb generation with tunable repetition rate, tunable lasing, and tunable processing and delay of optical pulses.**

## 1. Introduction

Microphotonic devices and circuits commonly consist of one or multiple connected basic elements, such as waveguides, couplers, and ring resonators [1,2]. In addition to the requirements of high fabrication precision and low losses [2,3], the *tunability* of these circuits and devices is of critical importance for a variety of applications [4,5]. While more complex tunable microphotonics circuits may be designed to enable wide class of programmable transformations of optical signals (see e.g., [1]), simple microdevices, such as standing along tunable three-dimensional microresonators, allow for unique functionalities not possible to achieve by other means. For a variety of applications, the tunability of spherical, toroidal, and bottle microresonators has been demonstrated using mechanical stretching, heating, and nonlinear light effects including those in monolithic and specially coated microresonators [6-10]. In most of these approaches, it is only possible to tune series of wavelength eigenvalues simultaneously without noticeable change in their separation.

However, for several applications, which include cavity QED [8,11,12], optomechanics [13,14], frequency microcomb generation [15,16], optical signal processing and delay [4,5,17], and lasing [18-21], it is critical to have microresonators with *tunable eigenwavelength separation*. For example, the latter allows the creation of optical frequency microcomb generators and microlasers with continuously tunable repetition rate and wavelength and to tune the microresonator eigenfrequency separation in resonance with the frequency of its mechanical oscillations. Considerable variation of the eigenwavelength separation commonly requires the variation of microresonator dimension and/or its refractive index parameters by the quantity comparable with their original values. One approach to solve this problem consists in using Fabry-Perot microresonators with tunable mirror separation which contain the optical materials under interest [12,21,22]. Additional flexibility of tuning can be achieved by employing Fabry-Perot microresonators with a liquid material inside [21] or translating a wedge-shaped solid optical material to vary its dimensions inside the Fabry-Perot microresonator [23].

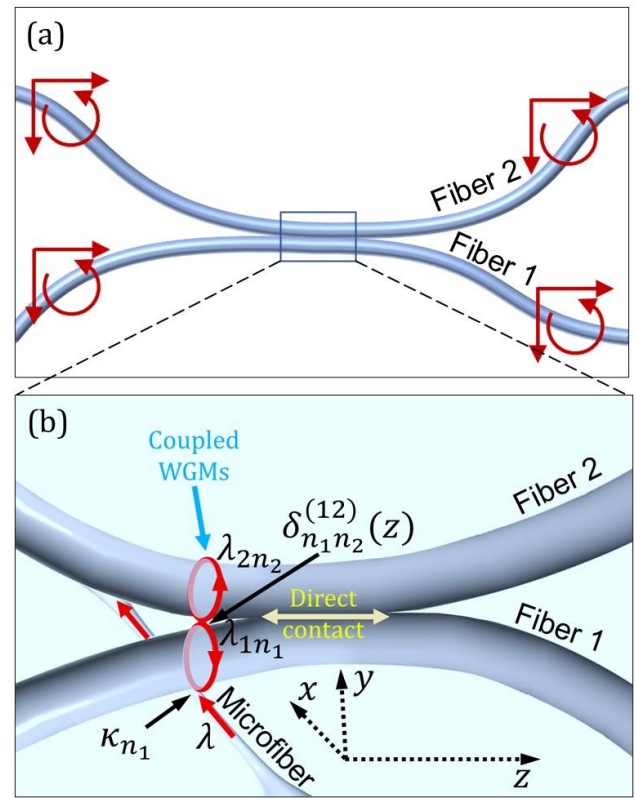

**Fig. 1.** (a) Coplanar bent optical fibers touching each other. The fiber profile is manipulated by bending and translation of the fiber tails indicated by curved and straight arrows. (b) Illustration of coupling between the input-output microfiber and WGMs in Fiber 1 and Fiber 2 near cutoff wavelengths.

Alternatively, of special interest is attaining the eigenwavelength separation tunability in three-dimensional monolithic high Q-factor microresonators, e.g., those with spherical, toroidal, and bottle shapes. This will allow us to add tunability to the emerging applications of these microresonators in QED, optomechanics, lasing, and frequency comb generation noted above. However, the deformation of most of these monolithic microresonators to achieve significant change of their eigenwavelength separation is unfeasible.

A unique exception, though, is exhibited by SNAP (Surface Nanoscale Axial Photonics) microresonators [24]. These microresonators are introduced at the surface of an optical fiber by its nanoscale deformation, which causes the nanoscale variation of the cutoff wavelengths (CWs) controlling the slow propagation of whispering gallery modes (WGMs) along the fiber axis (see [24,25] and references therein). In Ref. [26], a SNAP microresonator induced and fully reconfigurable by local heating of an optical fiber was demonstrated. In Ref. [27], it was shown that it is possible to create a SNAP microresonator and control its dimensions by local bending of an optical fiber. Both approaches allow for tuning of eigenwavelength separation of microresonators by the quantity comparable to or larger than its original value. However, in both approaches, the induced microresonator shapes had limited flexibility and their characteristic axial dimensions could not be reduced below several millimeters. In the first case, this restriction was caused by the imposed length of the characteristic heat distribution along the fiber. In the second case, the reduction of microresonator size was limited by the smallest curvature radius corresponding to the fiber breakage threshold.

In this paper we report on our discovery of a new type of WGM optical microresonators which belongs to the group of SNAP microresonators. We show that side coupled coplanar bent fibers (Fig. 1) can induce a high Q-factor SNAP microresonator localized in the region of fiber coupling. The configuration of fibers shown in Fig. 1 allows us to flexibly tune the shape of the induced SNAP microresonators and their axial dimensions from several tens of microns to several millimeters and, respectively, tune their eigenwavelength separation from hundreds of picometers to a picometer.

## Results

### Cutoff wavelengths of uncoupled and side-coupled straight fibers

First, it is instructive to consider the behavior of CWs for uncoupled and side-coupled *straight* optical fibers. For this purpose, we cleave a 125-micron diameter uncoated commercial optical fiber into two pieces (Fiber 1 and Fiber 2), which are then coaxially aligned and put into contact along 3.5 mm of their length as shown in Fig. 2(a). Light is launched into Fiber 1 by a transversely oriented taper with the micrometer diameter waist (input-output microfiber) connected to the Optical Spectrum Analyzer (OSA). After coupling into Fiber 1, light forms WGMs propagating along the fiber surface. In the region of direct contact of fibers (Fig. 2(a)), WGMs in Fiber 1 and Fiber 2 are coupled to each other.

To characterize the effect of interfiber coupling, we measured the spectrograms of the configured fiber system. For this purpose, the input-output microfiber was translated along Fiber 1 (Figs. 1(b) and 2(a)) touching it periodically with the spatial resolution of 2 μm. At the cut end of Fiber 1, the microfiber was moved towards Fiber 2 and continued scanning Fiber 2. The spectrograms of transmission power $P(\lambda, z)$ were measured as a function of wavelength $\lambda$ and microfiber position $z$ along the axis of Fiber 1.

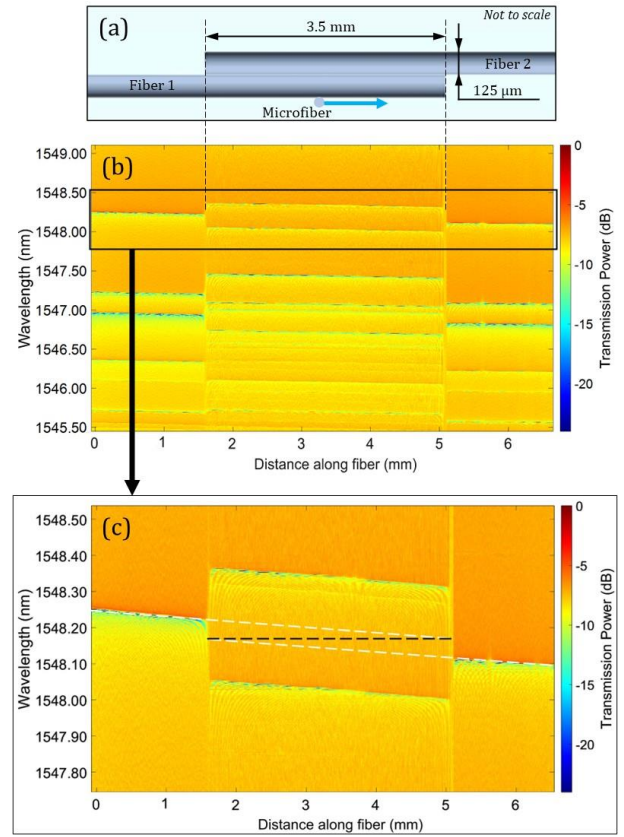

**Fig. 2.** (a) Illustration of side-coupled straight optical fiber configuration. (b) Spectrogram of this configuration. (c) Magnified section outlined in the spectrogram (b).

The measured spectrogram of our fiber system is shown in Fig. 2(b). The left- and right-hand sides of this spectrogram show the spectrograms of uncoupled Fiber 1 and Fiber 2, respectively. Lines in spectrogram shown in Fig. 2(b) indicate the CWs of uncoupled and coupled fibers. These CWs correspond to WGMs with different azimuthal and radial quantum numbers. The magnified copy of the section outlined in Fig. 2(b) is shown in Fig. 2(c). It is seen that the CWs appear as straight lines slightly tilted with respect to the horizontal direction. From the measured magnitude of tilt, $\varepsilon_t = 0.015$ nm/mm, we determine the linear variation of the fiber radius $\Delta r_t = r_0 \varepsilon_t / \lambda_0 = 0.6$ nm/mm [28]. In the latter rescaling relation, we used $r_0 = 62.5$ μm and $\lambda_0 = 1.55$ nm. By linear extrapolation of CWs of Fiber 1 and Fiber 2 (dashed white lines), we confirm that, as expected, their positions (horizontal black dashed line) coincide at the cut ends of these fibers.

At the 3.5 mm long region of fiber touching, WGMs in Fiber 1 couple to WGMs in Fiber 2 and the corresponding CWs split. The structure and positions of CWs in the touching region depend on the magnitude of coupling and will be further discussed below. Here we note that the value of CW splitting found, e.g., from Fig. 2(c) is ~ 0.1 nm, which coincides with characteristic values of CW variation in SNAP microresonators [24,25]. In particular, the positive CW shift in the coupling region leads to the WGM localization and creation of a microresonator which can be tuned by changing the length of the side-coupled fiber segment. In our current experiment, the Q-factor of the induced SNAP resonator was poor due to the scattering of light at the imperfectly cleaved fiber ends, which, typically, ensure around 70% WGM reflectivity [29]. Nevertheless, we suggest that the demonstrated resonator can be directly used to create miniature broadly tunable optical delay lines generalizing our previous results based on the SNAP microresonators with fixed dimensions [30, 31]. Indeed, in these devices the WGM pulses complete only a single round trip along the fiber axis and therefore their attenuation at the fiber facets may reduce the output light power by around 50% only. We also suggest that, after feasible improvement, the Q-factor of these microresonators can be significantly improved as further discussed below.

**Basic experiment**

In our proof-of-concept experiments, we used 125-micron diameter uncoated commercial silica optical fibers touching each other as shown in Fig. 1(a). The ends of Fiber 1 and Fiber 2 were bent and translated to arrive at the required profile of these fibers near their coupling region illustrated in Fig. 1(b). The fibers used were either originally straight or preliminary softened in a flame and bent permanently. As described in the previous section, WGMs were launched into Fiber 1 by a transversely oriented microfiber connected to the OSA. If the separation between Fiber 1 and Fiber 2 is small enough, WGMs penetrate from Fiber 1 into Fiber 2.

In the simplest configuration considered in this Section, Fiber 1 was straight, and coplanar Fiber 2 was bent. The fibers were put in contact and then slightly pushed towards each other to increase the coupling region. The photograph of the fiber configuration used in this experiment is shown in Fig. 3(a). From this picture, we estimated the curvature radius of the bent fiber as $R \sim 30$ mm (see further discussion of the fiber profile below). Fig. 3(b) shows the spectrogram of the configured structure measured along the 3.5 nm bandwidth within the 700 μm axial length of Fiber 1. At the edges of the scanned region, the interfiber coupling is negligible. In these regions, CWs do not noticeably change with distance $z$ and, thus, correspond to Fiber 1 only. The arrangement of CWs in these regions is similar to that in Fig. 2(b).

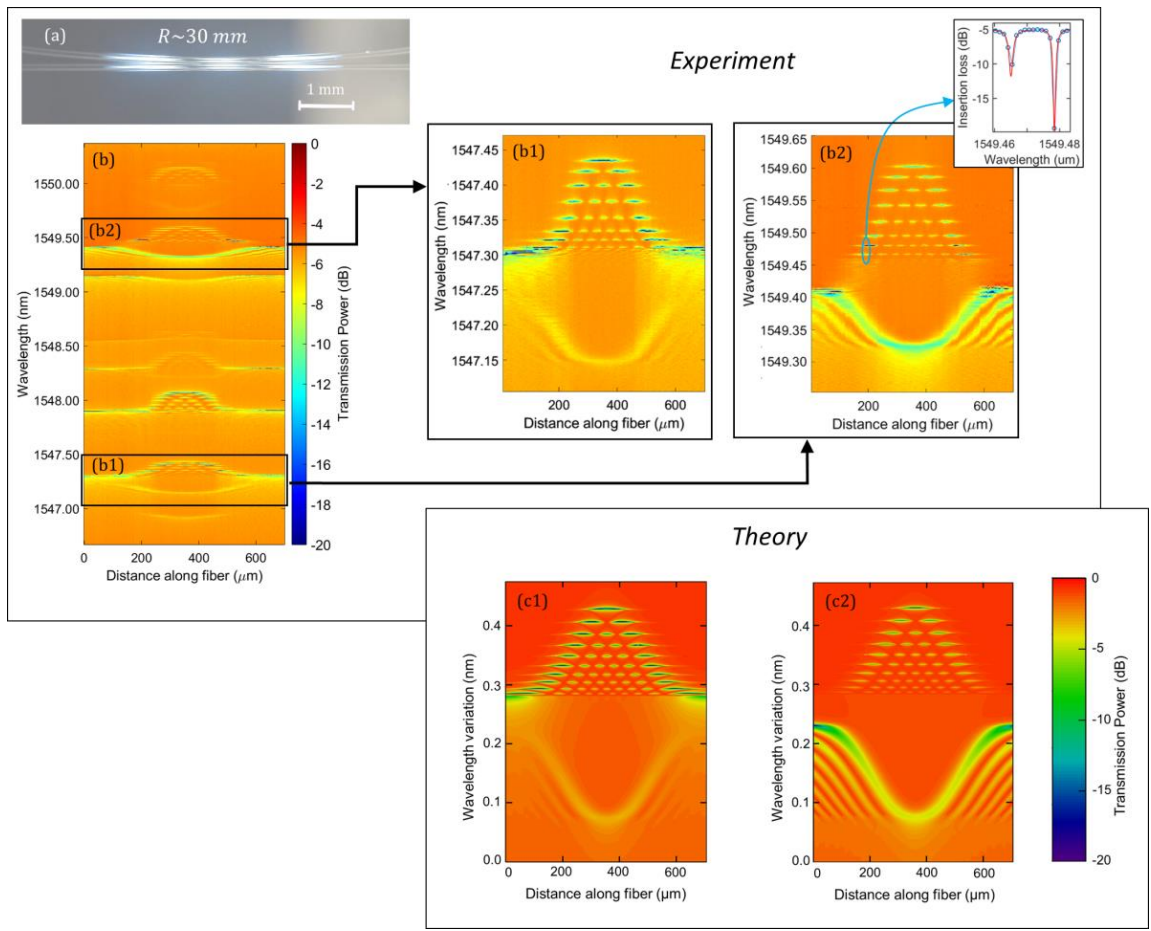

**Fig. 3.** (a) Photograph of the side-coupled fibers used in the experiment. The upper fiber is bent with the curvature radius $R \sim 30$ mm and the lower fiber has the curvature radius greater than 1 m. (b) The spectrogram measured in the vicinity of the coupling region of these fibers. (b1) and (b2) Spectrograms showing the magnified sections outlined in the spectrogram (b). (c1) and (c2) Spectrograms of the microresonators numerically calculated in the two-mode approximation detailed in the text, which replicate the experimental spectrograms in Figs. (b1) and (b2), respectively.

The effect of coupling shows up in the central region of the spectrogram in Fig. 3(b). In this region, different CWs exhibit different positive and negative variations along the axial length $z$. The exemplary regions of this spectrogram named (b1) and (b2) are magnified in Figs. 3(b1) and 3(b2), respectively. It is seen that, as expected, in contrast to negative variations, positive CW variations lead to the WGM confinement and the creation of microresonators. Our estimates illustrated in the inset of Fig. 3(b2) show that the Q-factor of the created microresonator (which measurement was limited by the 1.3 pm resolution of the OSA used) exceeds $10^6$. The observed CW variations in Figs. 3(b1) and (b2) can be explained by the theory described below.

**Basic theory**

We assume that the fiber bending is small enough so that the propagation of light along the axial direction of side-coupled fibers (Fig. 1(b)) can be considered as propagation along a single waveguide with asymmetric cross-section including both fibers. The wavelengths of slow WGMs are close to the CWs $\lambda_n(z)$ of this compound waveguide. To determine the complex-valued CWs $\lambda_n(z)$, we introduce the original CWs $\lambda_{1n_1} + \frac{i}{2}\gamma_{1n_1}$ and $\lambda_{2n_2} + \frac{i}{2}\gamma_{2n_2}$ of unbent Fiber 1 and Fiber 2 with the imaginary parts determined primarily by material losses and scattering of light at the fiber surface. We assume that there are $N_1$ and $N_2$ cutoff wavelengths in Fibers 1 and Fiber 2, respectively, which contribute to the resonant transmission, so that $n_j = 1,2,\ldots,N_j, j = 1,2$. We refer to the integers $n$, $n_1$ and $n_2$ as to the transverse quantum numbers. Variation of $\lambda_n(z)$ is caused by bending of fibers [27] and, in our case, primarily by their coupling. In the absence of the input-output fiber, the CWs of our system, $\lambda = \lambda_n(z)$, $n = 1,2,\ldots,N_1 + N_2$, are determined as the roots of the determinant:

$$\det(\lambda \mathbf{I} - \mathbf{\Xi}(z)) = 0 \qquad (1)$$

Here $\mathbf{I}$ is the unitary $(N_1 + N_2) \times (N_1 + N_2)$ matrix and matrix

$$\mathbf{\Xi}(z) = \begin{pmatrix} \mathbf{\Lambda}_1 + \mathbf{\Delta}_1(z) & \mathbf{\Delta}_{12}(z) \\ \mathbf{\Delta}_{12}^\dagger(z) & \mathbf{\Lambda}_2 + \mathbf{\Delta}_2(z) \end{pmatrix}. \qquad (2)$$

The submatrices in Eq. (2) determine the original CWs of Fiber 1 and Fiber 2, $\mathbf{\Lambda}_j = \{\lambda_{jn_j} + \frac{i}{2}\gamma_{jn_j}\}$, couplings inside each of the fiber caused by bending, $\mathbf{\Delta}_j(z) = \{\delta_{m_jn_j}^{(j)}(z)\}$, and interfiber couplings $\mathbf{\Delta}_{12}(z) = \{\delta_{m_1n_2}^{(12)}(z)\}$, $m_j, n_j = 1,2,\ldots N_j$.

As in SNAP [24], dramatically small nanometer and sub-nanometer scale variations of CWs $\lambda_n(z)$ along the compound fiber waveguide can localize WGMs and induce an optical microresonator having eigenwavelengths $\lambda_{qn}$ with axial quantum numbers $q$. Due to the smooth and small CW variation and proximity of the localized WGM wavelengths $\lambda_{qn}$ to $\lambda_n(z)$, the corresponding eigenmode can be presented as $E_{qn}(x,y,z) = \Psi_{qn}(z)\Omega_n(x,y,z)$ where the transverse WGM distribution $\Omega_n(x,y,z)$ is calculated at the CW $\lambda_n(z)$ and depends on $z$ parametrically slow [32], and function $\Psi_{qn}(z)$ determines the axial dependence of the microresonator eigenmode amplitude and satisfies the one-dimensional wave equation [24]

$$\frac{d^2\Psi_n}{dz^2} + \beta_n^2(z,\lambda)\Psi_n = 0, \quad \beta_n(z,\lambda) = \frac{2^{3/2}\pi n_r}{\lambda_n^{3/2}}\sqrt{\lambda_n(z) - \lambda}. \qquad (3)$$

where $n_r$ is the refractive index of the fibers.

The coupling parameters $\kappa_{qn}(z)$ between WGM $E_{qn}(x,y,z)$ and the input-output wave in the microfiber is determined by their overlap integral. Commonly, the microfiber diameter is much smaller than the characteristic axial variation length of $E_{qn}(x,y,z)$. For this reason, similar to the analogous approximation in the SNAP platform [24, 33], the coupling parameters $\kappa_{qn}(z)$ are proportional to the values of $E_{qn}(x,y,z)$ at the axial coordinate $z$ of the input-output microfiber. Then, calculations based on the Mahaux-Weidenmüller theory [34-36] presented in Supplementary Material allowed us to express the transmission power $P(\lambda,z)$ through the input-output microfiber coupled to the considered fiber configuration (Fig. 1(b)) as

$$P(z,\lambda) = \left| \frac{1 + \sum_{n=1}^{N_1+N_2} D_n^*(z) G_n(z,z,\lambda)}{1 + \sum_{n=1}^{N_1+N_2} D_n(z) G_n(z,z,\lambda)} \right|^2. \qquad (4)$$

Here $G_n(z_1, z_2, \lambda)$ is the Green's function of Eq. (3). Eq. (4) generalizes the expression for the transmission power previously derived in Ref. [24]. As shown below, functions $D_n(z)$ can be expressed through and have characteristic values similar to the coupling D-parameters which were experimentally measured previously and typically have the real and imaginary parts $\sim 0.01$ μm$^{-1}$ [24, 33]. Close to the resonance condition, $\lambda = \lambda_{qn}$, for sufficiently small losses and coupling, and separated CWs $\lambda_n(z)$, only one Green's function with number $n$ contributes to the sums in Eq. (4). Then, Eq. (4) coincides with that previously derived in Ref. [24]. However, generally, the contribution of more than one term to the sums in Eq. (4) may be significant.

Before the detailed description of the spectrograms in Figs. 2(b) and 3(b), we note that the transmission power plots in these figures characterize the CWs of the coupled fiber system determined by Eq. (1) viewed by the input-output microfiber and, subsequently, OSA. Therefore, the CWs of Fiber 2, which are the solutions of Eq. (1) but uncoupled from Fiber 1 cannot be seen by the OSA. On the other hand, the number of CWs which can show up in the coupling region can be as many as $N_1 + N_2$, i.e., significantly greater than the number $N_1$ of visible uncoupled CWs of Fiber 1 (see Fig. 2(b) as an example).

To clarify the effect of coupling between WGMs in adjacent fibers, we consider the two-mode approximation, $N_1 = N_2 = 1$, assuming that the wavelength $\lambda$ of the input light is close to an unperturbed single WGM CW $\lambda_{11} + \frac{i}{2}\gamma$ of Fiber 1 and a single CW

$\lambda_{21} + \frac{i}{2}\gamma$ of Fiber 2 having the same imaginary part. Consequently, in Fig. 1(b) we now set $n_1 = n_2 = 1$. We neglect the effect of the CW variation due to the fiber bending [27], which is usually smaller than the effect of fiber coupling, setting $\delta_{11}^{(j)} = 0$. Then, the CWs $\lambda_1(z)$ and $\lambda_2(z)$ of the compound fiber are found from Eq. (1) as

$$\lambda_{1,2}(z) = \frac{1}{2}(\lambda_{11} + \lambda_{21}) + i\gamma \pm \sqrt{\frac{1}{4}(\lambda_{11} - \lambda_{21})^2 + \left(\delta_{11}^{(12)}(z)\right)^2} \quad (5)$$

The dependence on the transverse coordinates $x$ and $y$ (Fig. 1(b)) of the compound WGM corresponding to CWs $\lambda_j(z)$ can be calculated as follows. We introduce the unperturbed WGMs in Fiber 1 and 2 (considered unbent and uncoupled) calculated at their CWs $\lambda_{11}$ and $\lambda_{21}$ as $\Omega_1^{(1)}(x, y)$ and $\Omega_1^{(2)}(x, y)$. Then, in the two-mode approximation, the compound modes generated by weak coupling of modes $\Omega_1^{(1)}(x, y)$ and $\Omega_1^{(2)}(x, y)$ are determined as [37]

$$\Omega_1(x, y, z) = \cos(\alpha)\Omega_1^{(1)}(x, y) + \sin(\alpha)\Omega_1^{(2)}(x, y),$$
$$\Omega_2(x, y, z) = -\sin(\alpha)\Omega_1^{(1)}(x, y) + \cos(\alpha)\Omega_1^{(2)}(x, y), \quad (6)$$
$$\tan(2\alpha) = \frac{2\delta_{11}^{(12)}(z)}{\lambda_{11} - \lambda_{21}}.$$

Consequently, the coupling parameters to the microfiber entering Eq. (4) at coordinate $z$ are

$$D_{1,2}(z) = \frac{D}{2}\left(1 \pm \frac{\lambda_1 - \lambda_2}{\sqrt{(\lambda_1 - \lambda_2)^2 + 4\left(\delta_{11}^{(12)}(z)\right)^2}}\right), \quad (7)$$

where $D$ is the $z$-independent coupling parameter between the input-output microfiber and Fiber 1 (refs. 24, 33).

To map the bent fiber axial profile $h(z)$ to the CW envelope profiles of the induced microresonators, we have to determine the relation between $h(z)$ and coupling coefficient $\delta_{11}^{(12)}(z)$. Similar to calculations in Refs. [38, 39], for the smooth and small $h(z)$ considered here, we find

$$\delta_{11}^{(12)}(z) = \delta_0 \exp\left(-\frac{2\pi}{\lambda}\left(n_r^2 - 1\right)^{1/2} h(z)\right), \quad (8)$$

where $\delta_0$ is $z$-independent. Assuming the simplest profile of the bent fiber having the curvature radius $R$ as

$$h(z) = z^2/2R \quad (9)$$

for silica fibers with $n_r = 1.44$, we estimate the FWHM of $\delta_{11}^{(12)}(z)$ as $z_{FWHM} \sim 0.5(\lambda R)^{1/2}$. At $\lambda \sim 1.55$ μm and $R \sim 30$ mm of our experiment, we have $z_{FWHM} \sim 100$ μm. From Eqs. (5) and (8), we find that the FWHM of the CW, depending on the value of $\lambda_{11} - \lambda_{12}$, is between $z_{FWHM}$ and $2z_{FWHM}$ which is only in qualitative agreement with the microresonator FWHM $z_{FWHM} \sim 250$ μm found from experimental data in Figs. 3(b1) and (b2).

The results of our numerical modeling in the two-mode approximation considered based on Eqs. (3)-(9) are shown in Figs. 3(c1) and 3(c2). To fit the experimental data, we set the average CW $0.5(\lambda_{11} + \lambda_{12}) = 1.55$ μm, the CW difference $\lambda_{11} - \lambda_{12} = 0.05$ nm in Fig. 3(c1) and $\lambda_{11} - \lambda_{12} = -0.05$ nm in Fig. 3(c2), coupling parameter $D = -0.01 + 0.01i$ μm$^{-1}$ (refs. 24, 33), Q-factor $Q = 10^6$, the microresonator FWHM $z_{FWHM} \sim 250$ μm and its spectral height $\sim 0.15$ nm, similar to these values found from Figs. 2(b1) and (b2).

The experimental spectrograms in Fig. 3(b1) and (b2) and theoretical spectrograms in Figs. 3(c1) and (c2) look nicely similar. However, important differences between them should be noted. From Eqs. (8) and (9), the FWHM value $z_{FWHM} \sim 250$ μm corresponds to the Fiber 2 curvature radius $R \sim 66$ mm, which is twice as large as that measured from the fiber image shown in Fig. 3(a). We suggest that the difference is caused by the deviation of the shape of Fiber 2 from parabolic in the coupling region as well as by the fiber misalignment. The additional deformation of fibers may be induced by their electrostatic attraction and pressuring, which are not visible in Fig. 3(a). Our suggestion is confirmed by the experimental profiles of the induced microresonator envelopes and CW shapes in Figs. 3(b1) and (b2) which, as compared to those in the theoretical spectrograms in Figs. 3(c1) and (c2), have larger side slopes and are flatter in the middle. Next, we notice that, in the theoretical spectrograms, the CW wavelength profiles are more mirror-symmetric to the microresonator envelopes with respect to the horizontal line (following Eq. (5)), while, in the experimental spectrograms, the lower CW profiles are shallower than the microresonator envelopes. We suggest that this deviation can be eliminated by taking into account the coupling with other WGMs ignored in the two-mode approximation considered.

**Tunability**

Bending and translating the tails of Fiber 1 and Fiber 2 side-coupled to each other as illustrated in Fig. 1 allowed us to tune the dimensions of the fiber coupling region and thereby tune the dimensions of created microresonators. As in the previous experiments, we used 125 μm optical fibers. We investigated the cases of the smallest microresonators containing a few wavelength eigenvalues and having the characteristic axial dimensions of hundred microns (Figs. 5(a1)-(a4)), as well as larger microresonators with dimensions of several hundred microns (Figs. 5(b1)-(b4) and (c1)-(c3)) and the largest microresonator having the axial length of 5 millimeters (Fig. 4(d)).

Considering the smallest microresonators, we monitored the process of their creation. Side-coupling of a straight Fiber 1 and Fiber 2 bent with a sufficiently small curvature radius of $\sim 1$ mm introduced small perturbation in CWs shown in the spectrogram in Fig. 4(a1). Increasing the fiber radius further, we arrived at the microresonator with a single eigenwavelength (Fig. 4(a2)). The inset inside the Fig. 4(a2) spectrogram, which magnifies the region near this eigenwavelength, shows that the axial dimension of the corresponding eigenmode is $\sim 200$ μm. Remarkably, except for the axial dimension of localized WGMs with uniform magnitude in specially designed bat microresonators [39, 40], this dimension (which expansion is critical, e.g., for QED applications [41]) is the record large characteristic WGM dimension demonstrated in microresonators to date. The measured Q-factor of this microresonator (limited by the 1.3 pm resolution of the OSA used) was slightly greater than $10^6$.

Larger bending radii of Fiber 2 having the order of 10 mm led to the creation of microresonators with millimeter-order axial dimensions having the spectrograms shown in Figs. 5(b1)-(b4) and (c1)-(c3). The close to parabolic shape of these microresonators suggests that they can be used, e.g., as tunable optical frequency comb generators [42]. We note that the behavior of the CWs and microresonators envelopes in most of these spectrograms cannot be accurately described by the two-mode approximation considered above. Of particular interest is the spectrogram shown in Fig. 4(c2). At first sight, the envelop of the microresonator in this spectrogram is the continuation of the CW of Fiber 1 (compare with Figs. 3(b1) and (c1)). Unexpectedly, the axial WGM localization in this microresonator (caused by the WGM reflection from the CW-generated turning points [24]) sharply dissolves inside the microresonator area.

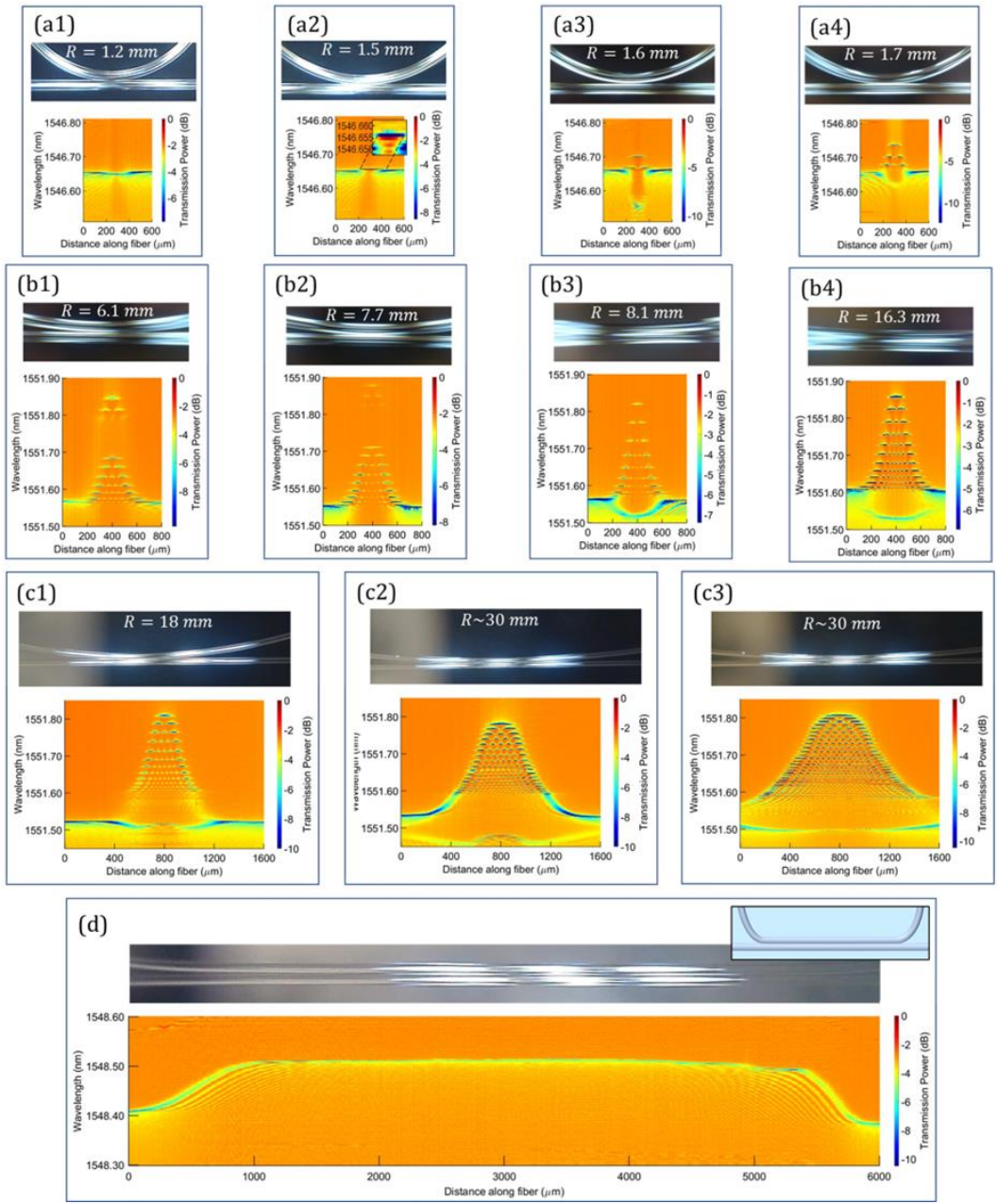

**Fig. 4.** Tunability of microresonators. (a1)-(a4) Spectrograms of induced microresonators for small curvature radius of Fiber 2 ~ 1 mm. (b1)-(b4) and (c1)-(c3) spectrograms of induced microresonators for a lager radius of Fiber 2 ~ 10 mm. (d) Spectrogram of a 5 mm long microresonator induced by touching straight Fiber 1 and Fiber 2 which was preliminary permanently bent at the ends as shown in the inset.

To create longer microresonators, we, first, permanently bent the tails of Fiber 2 as illustrated in the inset of Fig. 4(d). This allowed us to arrive at an arbitrarily large curvature radius of this fiber including its straight shape between the bent tails. As an example, Fig. 4(d) shows the spectrogram of a 5 mm long microresonator. Though the eigenwavelength width of this microresonator is greater than its free spectral range and its belong to the class of white light WGM resonators [43], we suggest that, in contrast to the lossy microresonators induced by side-coupled cleaved straight fibers (Fig. 2), its Q-factor is similar to that of the smaller microresonators with spectrograms shown in Figs. 3 and 5.

**Discussion**

The effect of induction of high Q-factor WGM tunable optical microresonators in side-coupled optical fibers discovered in this paper enables a range of exciting generalizations and applications. Further extension of tuning flexibility can be achieved by enabling different boundary conditions at the fiber tails (Fig. 1(a)), different interfiber touching stresses, and different preliminary permanent fiber bending.

Configurations of fibers, which are potentially attractive for future research and applications, are illustrated in Fig. 5. Fig. 5(a) shows a way to create long microresonators alternative to the method utilizing fibers with permanently bent tails illustrated in Fig. 4(d). In the configuration of Fig. 5(a), the length of the induced microresonator increases as the curvature radii of touching fibers approach each other. Provided that the variation of the fiber radii can be performed so that the parabolicity of the induced microresonators was maintained, the configuration of Fig. 5(a) can serve for the generation of the optical frequency combs with a tunable repetition rate.

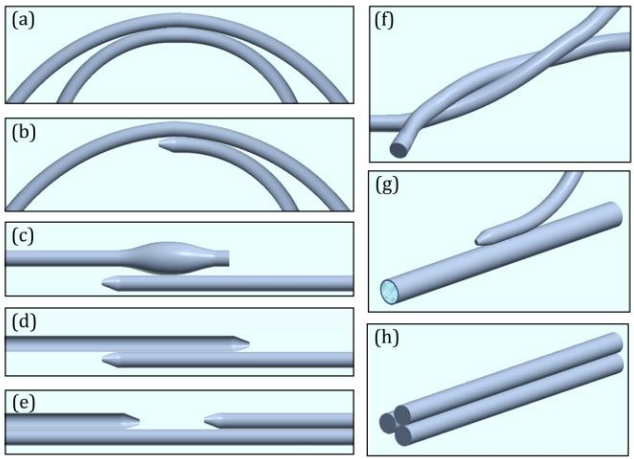

**Fig. 5.** (a) Bent fibers with increased coupling region. (b) Bent fibers with increased coupling region and abrupt side of the induced microresonator. (c) A bottle microresonator side-coupled to a fiber. (d) Side coupled straight fibers with tapered facets forming a rectangular microresonator. (e) Two straight fibers with tapered facets coupled to the third straight fiber forming a rectangular microresonator. (f) Twisted side-coupled fibers. (g) A microcapillary fiber filled with liquid and side coupled to a bent fiber. (h) Three straight coupled fibers.

In Fig. 5(b), the lower fiber is terminated with a short taper, which can be introduced using, e.g., a $CO_2$ laser. Simple estimates show that a taper with a characteristic length of 100 μm at the end of a 125 μm diameter optical fiber creates an abrupt CW barrier with a slope of ~ 100 nm/μm at 1.5 μm wavelength. The steepness of the slope of this barrier (critical for impedance matching of light from the input-output microfiber [44]) is 100 times greater than that demonstrated in Ref. [45] with the femtosecond laser inscription. The configuration shown in Fig. 5(b) can be used for the creation of miniature dispersionless tunable optical delay lines provided that the shape of the induced microresonator is kept semi-parabolic in the process of tuning [44].

Experimental investigation and development of the theory of WGMs in a microresonator side-coupled to an optical fiber is of particular interest. Fig. 5(c) illustrates the side coupling of a fiber and a bottle microresonator. While the fiber is open-ended, coupling of the bottle microresonator to the straight fiber can cause the localization of light in the fiber, similar to the coupling between bent optical fibers considered above. The configuration shown in Fig. 5(c) suggests a way of tuning the microresonator eigenwavelengths.

The fiber configuration shown in Fig. 5(d) is similar to two straight side-coupled fibers with cleaved ends (Fig. 2). To improve the Q-factor of the microresonator induced along the coupling region, the cleaved ends of fibers shown in Fig. 2(a) are modified by the tapered ends. The configuration of fibers shown in Fig. 5(e) illustrates an alternative way to create tunable microresonators when the position of both their sides can be tuned. The rectangular microresonators induced in both configurations can be used for the creation of tunable delay lines which, as shown in Ref. [31], can be dispersionless with a good accuracy.

The coupling of twisted optical fibers illustrated in Fig. 5(f) is interesting to investigate both theoretically and experimentally. In the cylindrical coordinates $(z, \rho, \varphi)$ of one of the fibers, the curve along which the fibers touch each other corresponds to the azimuthal angle $\varphi = \varphi_0 + \alpha z$, where $\alpha$ is the twisting coefficient. The corresponding value of the WGM field is proportional to $\exp[i\beta z + im(\varphi_0 + \alpha z)]$ where $\beta$ is the propagation constant. From this expression, a WGM at CW corresponding to $\beta = 0$ is seen by another fiber as a mode with nonzero propagation constant. Thus, in contrast to the untwisted fibers, coupling between the side-coupled twisted fibers is essentially three dimensional.

Fig. 5(g) shows a microcapillary fiber filled with liquid and side-coupled to a bent fiber. For the microcapillary with sufficiently thin walls, a microresonator induced inside it by the side-coupled fiber performs nonlocal sensing of liquid [46]. In Refs. [47, 48], such microresonators were introduced with the $CO_2$ laser and slow cooking methods. Fig. 5(g) suggests the simplest approach for the realization of nonlocal microfluidic sensing.

Fig. 5(h) illustrates three straight side-coupled fibers. In contrast to two coupled fibers, WGMs launched into this configuration will propagate into both azimuthal direction and, in particular, into the positive and negative directions of the input-output microfiber with approximately the same amplitudes. The channel formed between these fibers can be used for gas and microfluidic sensing. Unlike the microcapillary illustrated in Fig. 5(g), no ultrathin wall enabling the WGM sensing of the internal channel is required in this case.

While the model of two coupled CWs developed here qualitatively explains some characteristic features of the experimentally measured spectrograms, the complete explanation and quantitative fitting of the experimental data should include the effect of several CWs and be based on the further development of the coupled wave theory. The future theory should also allow us to express the fiber profiles and deformation in the region of coupling through the values of forces and moments applied to the fiber tails (Fig.1(a)) including the effect of electrostatic fiber attraction.

We suggest that the fixed submicron-wide gaps between coupled fibers and input-output microfiber, rather than their direct contact considered here, will allow us to demonstrate the proposed microresonators with the Q-factor exceeding $10^8$ (ref 8). While such large Q-factors are not required for the realization of tunable delay lines [44], signal processors [25], and microlasers [19-21], they may be important for the realization of frequency comb generators with tunable repetition rate [15,16,42], as well as for the cavity QED [8,11,12] and optomechanical applications [13,14].

## Supplementary material

### Expression for the transmission power

We introduce the discrete eigenwavelengths of the microresonator in the compound fiber system, $\lambda_m + \frac{i}{2}\gamma_m$, $m = 1, 2, \ldots, M$ and coupling coefficients $\kappa_m(z)$ between the corresponding eigenmodes and the input-output microfiber positioned at axial coordinate $z$. We calculate the transmission power $P(\lambda, z)$ of our system by applying the Mahaux-Weidenmüller formula [34-36]:

$$P(\lambda, z) = |1 + iT(\lambda, z)|^2,$$
$$T(\lambda, z) = \mathbf{K}^\dagger(z)\left(\Delta(\lambda) - \tfrac{i}{2}\mathbf{K}(z)\mathbf{K}^\dagger(z)\right)^{-1}\mathbf{K}(z) \quad (S1)$$

where

$$\Theta(z) = \Delta(\lambda) + \tfrac{i}{2}\mathbf{K}(z)\mathbf{K}(z)^\dagger, \quad \mathbf{K}(z) = \begin{pmatrix} \kappa_1(z) \\ \kappa_2(z) \\ \ldots \\ \kappa_M(z) \end{pmatrix},$$

$$\Delta(\lambda) = \begin{pmatrix} \lambda - \lambda_1 - \tfrac{i}{2}\gamma_1 & 0 & \ldots & 0 \\ 0 & \lambda - \lambda_1 - \tfrac{i}{2}\gamma_1 & \ldots & 0 \\ \ldots & \ldots & \ldots & \ldots \\ 0 & 0 & \ldots & \lambda - \lambda_M - \tfrac{i}{2}\gamma_M \end{pmatrix}. \quad (S2)$$

It is assumed in Eq. (S1) that the coupling to the input-output waveguide does not introduce the shifts of the eigenwavelengths [35] which will be added later. We simplify the expression for the transmission power by expanding the inverse matrix in Eq. (S1) as follows:

$$\left(\Delta(\lambda) - \tfrac{i}{2}\mathbf{K}(z)\mathbf{K}^\dagger(z)\right)^{-1}$$
$$= \sum_{n=0}^{\infty} \left(\tfrac{i}{2}\right)^n \left(\Delta(\lambda)^{-1}\mathbf{K}(z)\mathbf{K}^\dagger(z)\right)^n \Delta(\lambda)^{-1}$$
$$= \Big[1 + \tfrac{i}{2}\Delta(\lambda)^{-1}\mathbf{K}(z)\mathbf{K}^\dagger(z)$$
$$+ \left(\tfrac{i}{2}\right)^2 \Delta(\lambda)^{-1}\mathbf{K}(z)\mathbf{K}^\dagger(z)\Delta(\lambda)^{-1}\mathbf{K}(z)\mathbf{K}^\dagger(z) +$$
$$+ \ldots + \left(\tfrac{i}{2}\right)^n \Delta(\lambda)^{-1}\mathbf{K}(z)\left(\mathbf{K}^\dagger(z)\Delta(\lambda)^{-1}\mathbf{K}(z)\right)^{n-1}\mathbf{K}^\dagger(z) + \ldots \Big]\Delta(\lambda)^{-1}$$
$$= \left[1 + \tfrac{i}{2}\Delta(\lambda)^{-1}\mathbf{K}(z)\mathbf{K}^\dagger(z)\sum_{n=0}^{\infty}\left(\tfrac{i}{2}\right)^n\left(\sum_{m=1}^{M}\frac{|\kappa_m(z)|^2}{\lambda - \lambda_m - \tfrac{i}{2}\gamma_m}\right)^n\right]\Delta(\lambda)^{-1}$$
$$= \left(1 + \frac{\tfrac{i}{2}\Delta(\lambda)^{-1}\mathbf{K}(z)\mathbf{K}^\dagger(z)}{1 - \tfrac{i}{2}\sum_{m=1}^{M}\frac{|\kappa_m(z)|^2}{\lambda - \lambda_m - \tfrac{i}{2}\gamma_m}}\right)\Delta(\lambda)^{-1}$$

Substituting this expression into Eq. (S1), we find:

$$P(\lambda, z) = \left|\frac{1 + \tfrac{i}{2}\sum_{m=1}^{M}\frac{|\kappa_m(z)|^2}{\lambda - \lambda_m - \tfrac{i}{2}\gamma_m}}{1 - \tfrac{i}{2}\sum_{m=1}^{M}\frac{|\kappa_m(z)|^2}{\lambda - \lambda_m - \tfrac{i}{2}\gamma_m}}\right|^2 \quad (S4)$$

We separate the series of eigenwavelengths $\lambda_m + \frac{i}{2}\gamma_m$ and coupling coefficients $\kappa_m(z)$ by their correspondence to CWs $\lambda_n(z)$ entering Eq. (3) of the main text. For this purpose, we rewrite these parameters as $\lambda_{qn} + \frac{i}{2}\gamma_n$ and $\kappa_{qn}(z)$, where $q$ is the axial quantum number of the eigenmode $E_{qn}(x, y, z) = \Psi_{qn}(z)\Omega_n(x, y, z)$. Here $\Psi_{qn}(z)$ satisfies Eq. (3) and $\Omega_n(x, y, z)$ is a parametrically slow function of the axial coordinate $z$. Substituting $\gamma_m \to \gamma_n$ we assume that the material losses do not depend on the axial quantum number $q$. Then, similar to the arguments of Ref. [24] (see Eq. (13) in this reference), the coupling coefficients can be factorized as $|\kappa_{qn}(z)|^2 = 2iD_n(z)|\Psi_{qn}(z)|^2$. Using the expression for the Green's function of Eq. (3),

$$G_n(z, z, \lambda) = \sum_q \frac{|\Psi_{qn}(z)|^2}{\lambda - \lambda_{qn} - \tfrac{i}{2}\gamma_n}, \quad (S5)$$

we rewrite Eq. (S4) as

$$P(\lambda, z) = \left|\frac{1 + \sum_{n=1}^{N} D_n^*(z) G_n(z, z, \lambda)}{1 + \sum_{n=1}^{N} D_n(z) G_n(z, z, \lambda)}\right|^2. \quad (S6)$$

To identify the physical meaning of parameters $D_n(z)$, we recall the expression for the transmission power of a SNAP microresonator under the assumption of a single CW contribution ($N = 1$) and

lossless coupling to the input-output microfiber[24]:

$$P_1(\lambda, z) = \left| \frac{1 + D_1^* G_1(z, z, \lambda)}{1 + D_1 G_1(z, z, \lambda)} \right|^2 \quad (S7)$$

Here complex parameter $D_1$, which was experimentally measured and analyzed previously [24, 33], determines the coupling to the input-output microfiber as well as the WGM phase shift due to this coupling. Importantly, while the imaginary part of $D_n(z)$ contributes to the widths of the resonances, its real part (not taken into account in the original Eq. (S1)) determines the WGM phase shifts caused by the coupling to the input-output microfiber.

**Funding.** The Engineering and Physical Sciences Research Council (EPSRC), grants EP/P006183/1 and EP/W002868/1. Horizon 2020 MSCA-ITN-EID grant 814147.

**Disclosures.** The authors declare no conflicts of interest.

**Data availability.** Data underlying the results presented in this paper are not publicly available at this time but may be obtained from the authors upon reasonable request.